# Balance of enthalpy and entropy in depletion forces

Shahar Sukenik[†], Liel Sapir[†] and Daniel Harries*

Institute of Chemistry and the Fritz Haber Research Center,
Hebrew University of Jerusalem, Israel

† These authors contributed equally.
* To whom correspondence should be addresses: daniel@fh.huji.ac.il

# Abstract

Solutes added to solutions often dramatically impact molecular processes ranging from the suspension or precipitation of colloids to biomolecular associations and protein folding. Here we revisit the origins of the effective attractive interactions that emerge between and within macromolecules immersed in solutions containing cosolutes that are preferentially excluded from the macromolecular interfaces. Until recently, these *depletion forces* were considered to be entropic in nature, resulting primarily from the tendency to increase the space available to the cosolute. However, recent experimental evidence indicates the existence of additional, energetically-dominated mechanisms. In this review we follow the emerging characteristics of these different mechanisms. By compiling a set of available thermodynamic data for processes ranging from protein folding to protein-protein interactions, we show that excluded cosolutes can act through two distinct mechanisms that correlate to a large extent with their molecular properties. For many polymers at low to moderate concentrations the steric interactions and molecular crowding effects dominate, and the mechanism is entropic. To contrast, for many small excluded solutes, such as naturally occurring osmolytes, the mechanism is dominated by favorable enthalpy, whereas the entropic contribution is typically unfavorable. We review the available models for these thermodynamic mechanisms, and comment on the need for new models that would be able to explain the full range of observed depletion forces.

# Highlights

- Depletion forces can drive association and compaction of macromolecules
- These forces appear upon the addition of preferentially excluded cosolutes
- Considered to stem from steric interactions and therefore are entropic in nature
- However, experimental data shows many cosolutes exert enthalpic depletion forces
- Cosolute chemical nature correlates with thermodynamic driving force

# 1. Introduction

Solvated colloids or macromolecules experience effective mutual attraction when the solution contains solutes that are depleted from the macromolecular interface. These *depletion interactions* promote compaction, association, or aggregation of the macromolecules. This review focuses on the molecular mechanisms that can govern this effective depletion "force". It has long been realized that steric interactions between the *preferentially excluded* solute (or "cosolute") and macromolecules play an important role in dictating the entropic nature of depletion attractions. More recently, however, the existence of enthalpically driven depletion forces has come to light.

Using a wide range of available experimental evidence, we review the range of possible depletion interactions. With this we attempt to provide an overview and general guidelines to mechanistically differentiate these important effective forces. We find that, at least where protein interfaces are involved, excluded cosolutes show a small number of distinct thermodynamic behaviors (or "fingerprints") that allow a simple yet practical categorization of molecular mechanisms. Importantly, the mechanistic behavior of a specific cosolute intimately links to the chemical family it belongs to.

Over a century of studies has taught us that the stability of macromolecules sensitively depends upon the composition of the bathing solution. For example in aqueous solutions, where many biologically relevant macromolecules act and interact, solvation of proteins depends not only on the concentration of added salt, but also on the specific ions that make up the salty solution. Hofmeister was first to note that anions such as bromide are less effective at precipitating proteins out of solution than fluoride[1,2].

Many proteins possess a native folded structure that carries specific functions in living cells. This structure is, however, typically only marginally stable in physiological solutions, and the structure can be lost under various deleterious solution conditions[3]. Certain solutes, such as urea and guanidinium salts, were found by Tanford and others[4,5] to shift the folding equilibrium towards the unfolded state. Other solutes like alcohols, detergents, as well as the pH and temperature of solution, can act in similar ways.

Here we shall primarily focus on cosolutes that are depleted, and hence preferentially excluded from macromolecular interfaces, such as macromolecular crowders or osmolytes. These excluded cosolutes have been used by scientists for decades to stabilize proteins in their folded state in solution or to exert effective attractive interactions between colloids and macromolecules[6–10]. The same strategy has been used by nature to keep cellular macromolecules stable even under harsh deleterious environmental conditions in perpetual proteostasis[11–13]. Depletion interaction not only shifts equilibrium towards the more compact, native state of proteins but can also drive protein oligomerization and precipitation[14–16]. Both these processes involve burial of exposed molecular surfaces, which becomes thermodynamically more favored by introducing excluded cosolutes. We shall, therefore, discuss both types of processes, bearing in mind this close analogy.

## 2. Cosolute-induced thermodynamic stabilization

From a thermodynamic perspective, Gibbs was probably first to explain the action of solute addition to solutions. He considered an extended interface, like that of a large colloid, bathed in a solution containing solvent and an added solute or "cosolute" at concentration $c_c$. If this cosolute is attracted to the interface, then at equilibrium there will be a net excess number of the cosolute, $\Gamma_c$ (or deficit of solvent, e.g. water, $\Gamma_w$) at the interface with respect to its concentration in the bulk of the bathing solution. In the presence of this cosolute that is "preferentially included", the interface is necessarily stabilized (its free energy is lowered) with respect to the same interface bathing in the solvent in the absence of cosolute. Conversely, if cosolute is excluded from the interface, that interface is destabilized and a cosolute deficit is formed at the interface with respect to the bulk.

The "Gibbs adsorption isotherm" embodies this idea, by directly relating the excess in free energy per interface unit area (or surface tension) $\Delta G_{surf}$ with the surface excess or preferential inclusion of cosolute and solvent at the interface, $-d\Delta G_{surf} = \Gamma_c d\mu_c + \Gamma_w d\mu_w$, where $c$ and $w$ subscripts refer to cosolute and water, respectively, and $\mu$ represents the chemical potential. For the limiting case where cosolute concentrations are low, this simply means that the change in free energy due to cosolute addition is related to the excess or deficit of water or cosolute at the interface: $\partial \Delta G_{surf}/\partial c_c = (RT/N_w)\Gamma_w$ or equivalently, $-\partial \Delta G_{surf}/\partial \ln c_c = RT\Gamma_c$, where $c$ is on the molal scale and $N_w = 55.6$ mol kg$^{-1}$ is the number of molecules in a kilogram of water[17,18].

Similar expressions were subsequently derived to more directly address the action of various cosolutes or molecular ligands that interact with proteins, rather than extended interfaces. If in some process the extent of exposed macromolecular interface is reduced, this state will be favored in solutions of excluded cosolute, and disfavored in included cosolutes. Perhaps best known is the Wyman linkage[19,20] that relates equilibrium constants (such as folding or unfolding of a protein, or protein-protein interactions) with the changes in the deficit or excess of water or cosolute in the process. Notably, a rigorous discussion of the changes in solution structure due to cosolutes, and how these affect the macromolecular solvation were analyzed by Kirkwood and Buff[18,21–23].

The effect of excluded cosolutes can be followed, for example, in the folding process of a 16-amino acid peptide, met16, into a beta-hairpin structure, Figure 1. This peptide exists in equilibrium between a β-sheet and an unfolded ensemble, shown schematically in Fig. 1B, inset. In neutral pH buffered water at room temperature, $\Delta G_{folding} \approx -1$ kJ/mol for the folding process[24]. This makes the peptide particularly convenient for detection of the effect of cosolutes since its two-state folding process is relatively easily detected using circular dichroism. This allows the direct calculation of $\Delta G_{folding}$ even for small changes in solution conditions, without the need for the assumptions inherent in calculations from protein melting curves. In addition, the peptide is readily soluble under various conditions, and aggregation (in the form of amyloid fibrils) is observed only after several hours[25], a substantial time for experiments directed at the denatured (*D*) to native (*N*) equilibrium of the monomer. In this folding process, the free energy of folding $\Delta G_{folding} = G_N - G_D$ is related to $\Delta G_{surf}$, in that the extent of exposure to solution is reduced upon folding. Therefore, following the discussion above, solutes that are preferentially excluded from the peptide are expected to destabilize both the folded and unfolded states, but they should more weakly destabilize the more compact, folded state. The net effect is to shift equilibrium towards the peptide folded ensemble, and the change in the folding free energy due to cosolute measured by $\Delta\Delta G = \Delta G_{folding}^c - \Delta G_{folding}^w$ will decrease. This effect is indeed observed for molecularly small osmolytes, as well as for larger polymers such as polyethylene glycol (PEG), Fig. 1, and the same derived thermodynamic description can be applied to any process that involves the reduction of excluded volume, be it ligand binding, protein-protein interaction, or oligomerization.

For low enough cosolute (and protein) concentrations, the change in stability is directly related to the change in preferential hydration upon folding $\partial\Delta\Delta G / \partial c_c = (RT/N_w)\Delta\Gamma_w$. As seen for the example of met16 in Fig. 1 and in numerous additional examples, this change in preferential hydration upon folding is constant over a significant range of concentrations, and for many cosolutes. This also implies that the change in cosolute exclusion (i.e., the cosolute deficit with respect to the bulk) is linear in concentration, because $\Gamma_c = -(c_c/N_w)\Gamma_w$.

The notion that individual units of the macromolecular polymer can exclude cosolutes to varying extents has led to studies that effectively dissect the change in stabilization free energy into terms involving individual macromolecular components, such as peptide side-chains or specific chemical groups on proteins and peptides. Similar strategies have been successful in predicting the stabilizing (or destabilizing) effect of cosolutes based on the sum of contributions related to each dissected component and the extent of macromolecular exposure to solution[26–28].

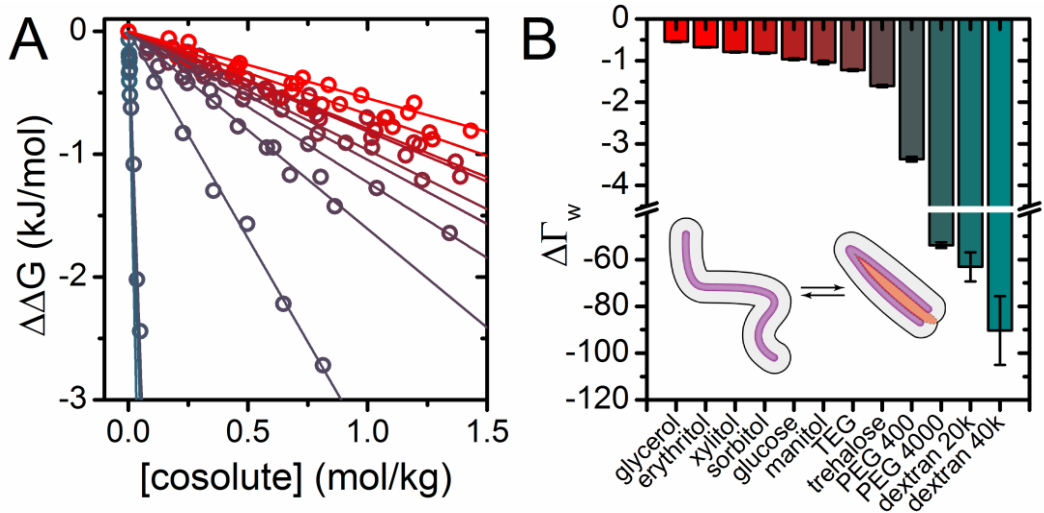

**Fig. 1. Cosolute impact on the folding of the met16 peptide.** (A) The change in free energy as a function of cosolute molality. Lines are linear fits of the data. Color represents cosolute identity, as detailed in panel B. (B) Change in the preferential hydration upon folding, $\Delta\Gamma_w$, for the different cosolutes shown in panel A. The cosolutes are arranged from the molecularly small (glycerol) to large (dextran), showing that $\Delta\Gamma_w$ tracks molecular size. Error bars represent the standard error of the linear fits. Inset: Scheme of the folding process of the 16 amino-acid met16 peptide (purple), with the surrounding grey region depicting the volume excluded to cosolute, and red highlighting the reduction of this excluded volume upon folding.

Interestingly, the extent of change in preferential hydration tracks the size of cosolute, Fig. 1B. The extent of preferential hydration can be thought of as approximately proportional to a volume from which cosolute is excluded, $\Delta V_{ex} \cong \Delta\Gamma_w v_w$, where $v_w$ is the molar volume of water. Therefore, the fact that changes in preferential hydration between the folded and unfolded state track the cosolute molecular size suggests that steric interactions may play an important role in the extent of cosolute preferential exclusion from the macromolecular interface. We return to this point in the next section, where we discuss the role of these excluded volume interactions in exerting entropically driven depletion forces that compact macromolecules or promote their associations.

## 3. Linking entropy-driven stabilization and steric interactions

Accumulating evidence, such as the dependence of the experimentally determined $\Delta\Delta G$ with concentration and the scaling with cosolute size, inspired Minton and coworkers to relate cosolute effects to steric interactions or "crowding" between the macromolecule and cosolute [29,30]. The importance of steric interactions in inducing depletion forces goes back to the work of Asakura and Oosawa (AO) [31,32]. Their model, initially developed for colloidal suspensions, was the first molecular-level theory that described the effective attraction between non-directly interacting colloidal particles immersed in a solution of excluded cosolutes.

In the AO model, the solvent is accounted for only implicitly, and the cosolutes interact with the colloid through hard-core interactions, thereby excluding them from a depletion layer around the colloid. Concomitant with colloid association, the cosolute-excluding volume decreases by $\Delta V_{ex}$. The added stabilization due to cosolutes, $\Delta\Delta G$, is therefore simply the reversible work associated with reducing the cosolute-excluding volume, hence $\Delta\Delta G = \Pi \Delta V_{ex}$, where $\Pi$ is the solution osmotic pressure. In the low concentration regime, where the van 't Hoff relation $\Pi = c_c RT$ applies, this translates to $\Delta\Delta G \sim c_c \Delta V_{ex}$. Since the AO model includes only hard-core interactions, the thermodynamic origin of the induced depletion forces is *necessarily purely entropic*, and depends only on the volume excluded to cosolutes and on cosolute concentration.

Consider, for example, a previously proposed model[33,34] for the association of two hard sphere "monomers" of radius $r_m$ in a solution of hard particles with radius $r_c$ into a spherocylinder "dimer" of equal total volume, Fig. 2A. This dimerization is accompanied by a decrease in excluded volume of

$$\Delta V_{ex}^{dim} = -\frac{4}{3}\pi r_c^3 \left(1 + \frac{r_m}{r_c}\right)^2.$$

For this process, the AO model predicts a linear dependence of $\Delta\Delta G$ on $c_c$, with a slope proportional to $\Delta V_{ex}^{dim}$, as plotted in Fig. 2B. For higher concentrations, a more appropriate approximation may also include the second virial coefficient for the osmotic pressure of hard-particles, $\Pi = c_c RT\left(1 + 16\pi c_c r_c^3/3\right)$, resulting in deviations from a linear dependence, Fig 2B.

An alternative, more precise approach to analyze this crowding uses scaled particle theory (SPT)[35–37] to approximate the reversible work required to form a cavity that accommodates the two monomers versus the dimer, in a solution of hard-particle crowders[33,34]. Using SPT, Minton and coworkers derived a dependence of the dimerization free energy gained with increasing cosolute concentration, Fig. 2B. Importantly, in the highly dilute regime, where $c_c \to 0$, all these approaches converge to the same variation of the free energy with concentration, so that $\partial\Delta\Delta G/\partial c_c \sim \Delta V_{ex}^{dim}$, as shown in Fig 2B, and detailed in the SI.

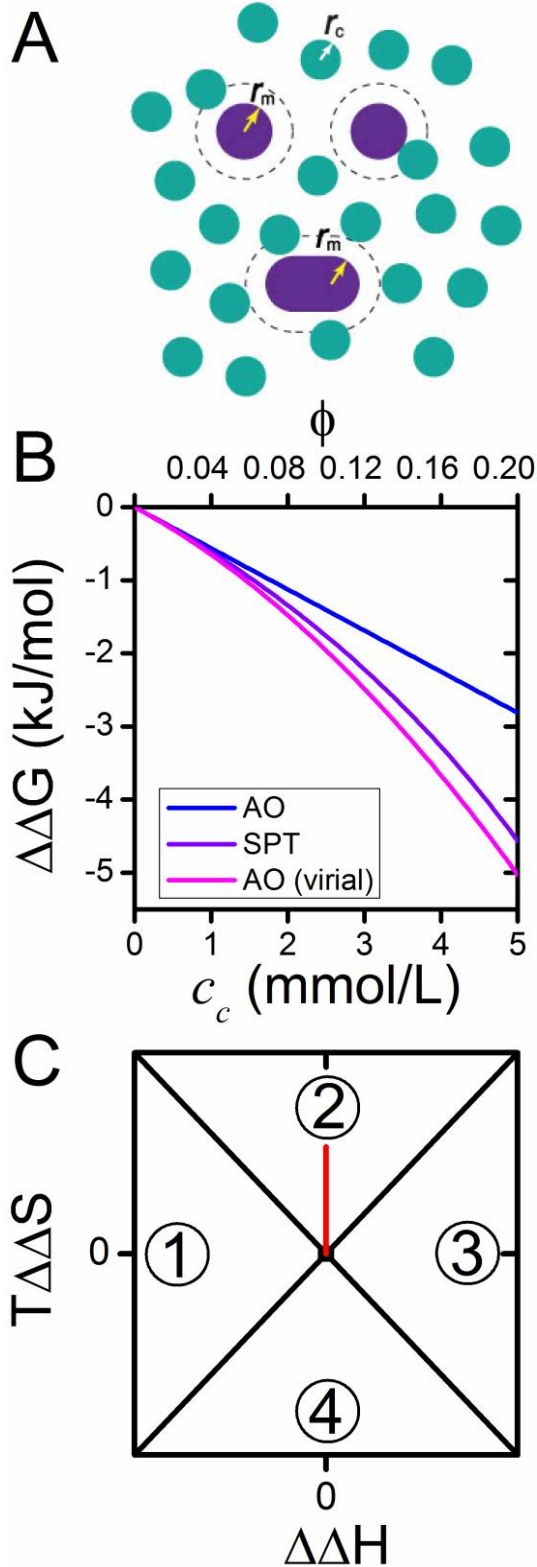

**Fig. 2. Entropic stabilization by cosolutes.** (A) Schematic of a dimerization process in the presence of hard excluded cosolutes. Crowders (cyan spheres) are excluded from the dashed volume, and thus exert a force stabilizing the dimer (purple spherocylinder) formation from two monomers (purple spheres). (B) Change in free energy of dimerization for the system in panel A with cosolute concentration. Here $r_m$= 3.5 nm and $r_c$ = 2.5 nm (the average radius for a 500 amino acid protein, and the radius of gyration of PEG4000 at 298K[75] , respectively). The change is calculated according to 3 models: AO (considering ideality of the cosolute) AO but with osmotic pressure including the second virial coefficient, and using SPT. Free energies are shown vs. cosolute concentration $c_c$ or volume fraction $\phi$. (C) enthalpy-entropy plot. Numbers represent sectors referred to in the text. Red line is the signature of entropic-driven stabilization by cosolutes.

## 4. Experimental evidences of entropic stabilization

To envisage excluded cosolutes as molecular "crowders" that drive macromolecular compaction through steric interactions necessarily invokes the idea of entropic stabilization. How accurate is this molecular-level picture? To address this question, it is instructive to follow different thermodynamic mechanisms through the corresponding enthalpic and entropic contributions, $\Delta\Delta H$ and $\Delta\Delta S$ [24,38]. These contributions are conveniently described by entropy-enthalpy plots that show the two contributions for different given cosolute concentration, Fig 2C. The diagonal $T\Delta\Delta S = \Delta\Delta H$ represents full compensation between entropy and enthalpy, and the area above this line implies a stabilizing cosolute effect. The diagonal $T\Delta\Delta S = -\Delta\Delta H$ further delineates the plot into four sectors. Sector 1 includes cosolutes exerting an enthalpically dominated stabilization, while sector 2 exhibits entropically dominated stabilization. Sectors 3 and 4 include destabilizing cosolutes with an enthalpic and entropic dominance, respectively. The extent of the cosolute effect on the free energy is described by the distance from the diagonal, which corresponds to the change in free energy as a result of cosolute stabilization, $\Delta\Delta G = \Delta\Delta H - T\Delta\Delta S$. Importantly, because they consider only hard-core repulsions, both AO and SPT models discussed in the previous section exhibit the exact same fully-entropic thermodynamic "fingerprint", as illustrated by the red line in sector 2, Fig. 2C.

Experimentally, the entropic and enthalpic contributions can be assessed directly through calorimetric measurements or indirectly through a van 't Hoff analysis. We follow experimental traces (or "fingerprints") of various excluded cosolutes in the entropy-enthalpy plane as they impact different protein processes. There have been numerous reports on the thermodynamic alteration to biomolecular stability as well as to protein-protein interactions induced by cosolute addition. To the best of our knowledge, however, the implications of the enthalpic and entropic contributions to these changes have not been previously analyzed and addressed in this context. We have, therefore, compiled several studies to resolve and follow the underlying thermodynamic mechanisms.

Most of the data extracted here has been measured using differential scanning calorimetry (DSC) or isothermal titration calorimetry (ITC). These methods have the benefit of providing thermodynamic values directly, without the need for van 't Hoff analysis, which could sometimes be prone to correlated errors in enthalpy and entropy[39]. Several works showed reversible melting curves from which we could reliably obtain the relevant enthalpy and entropy values. Taken together, the dataset encompasses diverse processes, including protein folding[38,40–46], protein-protein interactions[47] and ligand binding[48–50]. This allows a wide overview of biomolecular reactions for which the cosolute effect can be tested. Moreover, due to the large variety of studied proteins, cosolutes, and the processes they undergo, the measured changes in enthalpy and entropy encompass two orders of magnitude. The analyses performed on this collected data to extract the relevant thermodynamic quantities are described in the SI.

First we examine the effect for large polymeric crowders such as PEG, ficoll, and dextran. Although they are not naturally occurring, these crowders are often used to mimic the crowded cellular environment. These polymers have shown in many studies a qualitative and quantitative agreement with the entropic dominated SPT model both in folding and in more complex scenarios like amyloid aggregation [33,34,51]. For the met16 peptide, these cosolutes also showed the linear, cosolute-size dependent effect on folding, Fig. 1A.

From the accumulated information presented here, we find that the polymeric crowders (Fig. 3A) encompass both entropically and, to a lesser extent, enthalpically dominated stabilization mechanisms. The entropically dominated stabilization (sector 2), where the majority of the data is found, is in general agreement with models that describe entropic stabilization. All data points lie above the full compensation line, although associated free energy change is relatively small. A common trend seen in the data is that the entropic gain favoring folding is often mitigated by an enthalpic loss, in accordance with recent studies[33,52,53]. The origin of this unfavorable enthalpy has been associated with an effective attraction between cosolutes and macromolecules that acts in addition to the steric exclusion discussed in the previous section.

Several other data points, particularly for larger polymer concentrations, penetrate into the enthalpically stabilizing sector 1. While this is not directly predicted by AO theory, large polymers at even moderate concentrations can reach the so-called semi-dilute regime, where the idealization of the dilute solution of hard spheres is no longer relevant[54,55]. In these cases, enthalpically dominated "chemical" interactions likely arise, as we further discuss below. In fact, a depletion force that is purely entropic with no enthalpic component (favorable or unfavorable) is quite rare. A compelling yet non-molecular example is found in vibrofluidized granular materials where fully entropic depletion forces have been found to drive hard particle aggregation[56].

Interestingly, triethylene-glycol (TEG, composed of 3 ethylene glycol monomers, red symbols in Fig. 3A) is found in the entropically dominated sector. This cosolute is roughly the size of sorbitol (partial molar volume of ~130 and ~120 ml/mol for TEG and sorbitol in solution, respectively[25]), yet stabilizes peptide folding through a different thermodynamic mechanism. Whereas sorbitol acts through increasing the enthalpic gain (as we discuss in the next section), TEG stabilizes the peptide by favorable entropy that overcompensates an unfavorable enthalpy. These different thermodynamic mechanisms for equal-sized cosolutes highlight the importance of the chemical nature of the cosolute that goes beyond excluded volume. In addition, it may offer first insights as to the disparities observed between osmolytes and other crowders in complex processes such as peptide or protein aggregation into amyloids[14,25].

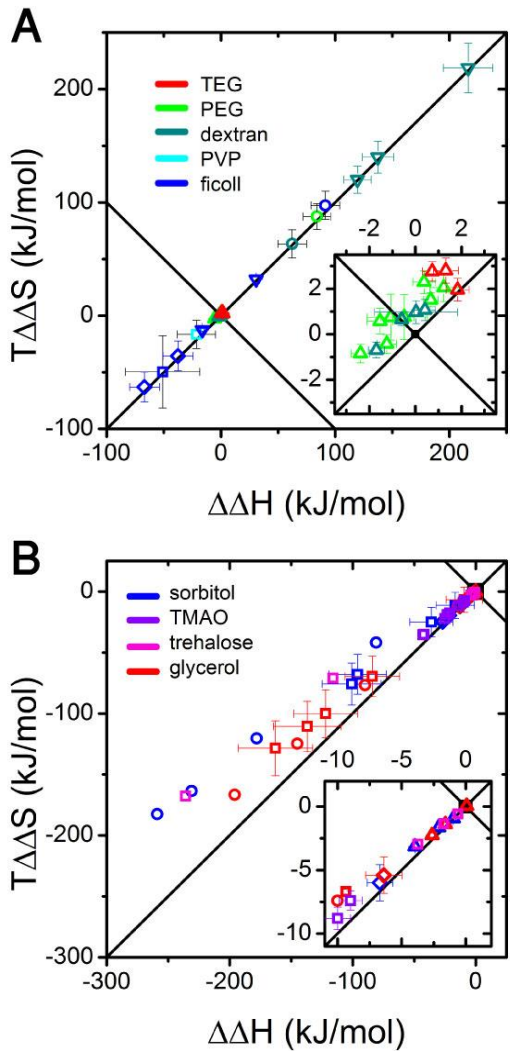

**Fig. 3. Thermodynamic fingerprints of cosolute effects on various protein processes.** Colors represent cosolute identity and symbol shapes represent the studied protein and process. (A) Entropy-enthalpy plot for polymeric crowders. Data points correspond to met16 folding (△)[38], SOD binding of catalase (○)[50], ubiquitin folding (□)[45], CI2 folding (◇)[40], and α-lactalbumin folding (▽)[46]. Inset: an enlarged part of the plot around the origin. (B) Entropy-enthalpy plot for osmolytes. Data points correspond to α-chymotrypsin folding (○)[43], met16 folding (△)[38], integrase DNA binding (◇)[48], and lyzosyme folding (□)[42]. Inset: an enlarged part of the plot around the origin. Error bars represent reported errors where available. The analysis of literature data is detailed in the SI.

## 5. Enthalpy-driven cosolute stabilization

The dissection of $\Delta\Delta G$ into its enthalpic and entropic components permitted first hints that not all cosolutes conform to an entropically driven mechanism. We have found that many cosolutes show stabilization that is enthalpically dominated. Even more surprising was the fact that entropy was actually destabilizing for the process, and was offset by favorable enthalpy, as shown in Fig. 3B. This trend was found for molecularly small excluded cosolutes, and specifically for a class of cosolutes termed osmolytes, Fig. 3B. These cosolutes are naturally occurring, and are used by cells to counteract external osmotic pressures, as well as other deleterious conditions. The enthalpically stabilizing mechanism persists even at high (>1M) concentrations for all cosolutes shown. This is found in several, very different experimental systems, indicating that the importance of this mechanism extends beyond a specific peptide or protein and is not limited to the process of protein folding. Furthermore, we notice that glycerol and TMAO, relatively small osmolytes, lie along a line whose slope is smaller than that of the other, larger osmolytes such as sorbitol and trehalose.

Taken together, these results contrast mechanisms that rely on direct steric repulsion between macromolecule and cosolutes, and the enthalpic dominated mechanism that is revealed cannot be explained by invoking excluded volumes alone. These findings not only suggest that at least two different mechanisms exist for cosolute stabilization, but that these mechanisms are closely correlated with the chemical identity of the cosolute: polymers typically act as entropic crowders (at least at low enough concentrations) while osmolytes act enthalpically, even in cases where their sizes are similar to a polymeric counterpart (as in the case of sorbitol and TEG).

## 6. Interactions that can drive enthalpic stabilization by cosolutes

Learning from the large number of examples we have amassed, we propose that the thermodynamic fingerprint of cosolute stabilization is related to the molecular mechanism by which these cosolutes act. Though there has been some controversy concerning the validity of the dissection of free energy into enthalpic and entropic terms[57–59], and the way these should be interpreted, we believe that the accumulated evidences over the wide gamut of macromolecular processes, proteins, and cosolutes uncover fundamental and ubiquitous mechanisms for cosolute effects. Additional confidence in the existence of the enthalpically dominated effect stems from the fact that we consider the net enthalpy and entropy of the perturbation caused by cosolute addition by subtracting the "background" of the protein process (i.e., subtracting the enthalpy and entropy of the process in the absence of cosolutes). This has been shown to give credible enthalpic and entropic contributions[39].

How can a favorable enthalpic depletion contribution, accompanied by unfavorable entropy, be explained? One aspect absent from models that address solutions of hard spheres is a direct and explicit consideration of the solvent, in our case water. Recall that in the AO and other steric exclusion models, the solvent is implicit or "transparent" (Fig. 2A), and plays no direct role in the forces exerted on the protein by the solution. In reality, it is possible that in the enthalpically dominated regime, the exclusion and depletion force is indirect, and is mediated by the solvent.

A possible explanation for the source of enthalpic contributions emerges when considering the macromolecule and surrounding solution. It has been established that the entropic and enthalpic contributions to $\Delta\Delta G$ can be further dissected into two components; the first includes all macromolecule interactions (both with itself and with the solution, i.e., solvent and cosolute, where solvent is the solution component in excess, usually water), and the second includes solution interactions only (solvent-solvent, cosolute-cosolute, and solvent-cosolute). The "entropy-enthalpy compensation theory" asserts that, though the solution-related enthalpic component could be large, it is completely balanced by a matching solution entropic component. Therefore, the net exerted stabilization $\Delta\Delta G$, represented by the distance from the full compensation diagonal in the enthalpy-entropy plot, is only determined by the macromolecule-related terms. Notwithstanding, the solution term can affect the total values of the enthalpic and entropic contributions, effectively shifting the cosolute contribution along a parallel line to the full compensation line in the entropy-enthalpy plot[60–62].

This dissection can help bridge the enthalpic-induced cosolute stabilization observed in experiments with the AO model. The macromolecule-related terms, which are the only ones accounted for explicitly in AO and SPT, could potentially show the behavior predicted for steric excluded cosolutes. In these models the sum of solute-related interactions are only implicitly considered by assuming that they result in the hard-core cosolute-macromolecule repulsion. Realistically, however, the effective macromolecule-cosolute repulsion may be largely affected by all intermolecular interactions, including those of the solvent. In concert, the solution terms (that do not directly alter $\Delta\Delta G$) could act to shift the enthalpic contribution to sector 1, making the stabilization appear enthalpically dominated. Unfortunately, there is no current experimental method to separately measure the macromolecule and solution terms directly.

A complimentary perspective addresses the explicit cosolute-induced changes occurring in the solution (cosolute and solvent mixture) itself. The existence of water that is structured differently at the macromolecular interface than in the bulk has been frequently found experimentally and in simulations for many different solutions of macromolecules, including, for example, osmolytes[63,64], lipids[65,66], proteins[67,68] and nucleic acids[69]. The "hydration force"[70] typically associated with this different water structuring decays exponentially with a typical decay length of roughly the dimensions of a 1-2 water molecules (2-4 Å) [71–73], and is insensitive to macromolecule identity. As a result of this different structuring, cosolutes are excluded since they are better solvated in the bulk. Thus, a high concentration of (almost pure) water is found in the first few layers of molecules surrounding the macromolecular surface. Simulations show water is isolated from the bulk by the first layer of cosolutes, and may rearrange in order to optimize its interactions (or hydrogen bonding network, in the case of water), resulting in structures somewhat different from their structure in the bulk[67]. This essentially creates a new solvation environment for the macromolecules, which depends on the interactions between water and the water-sequestering cosolute.

The change in water structure around the macromolecule could entail a change in enthalpy and can also explain the associated unfavorable entropy due to changes in water ordering at the interface. Sector 1 in Fig. 4 depicts a schematic of the proposed mechanism involving isolation of a hydration layer from bulk water (small blue circles). This water sequestration induces subsequent rearrangement to form water that is ordered differently than in bulk solution. This new ordering generates an entropic loss, concomitant with an enthalpic gain for the folding process for cosolutes found in sector 1.

While this model is hypothetical, it finds several supporting evidences. First, molecular dynamic simulations of the met16 peptide in the presence of sorbitol[67] show a change in the hydrogen bonds and rotational degrees of freedom for the sequestered water. In fact, both simulations and experiments show that water-water hydrogen bonds in the presence of polyol cosolutes were found to become stronger, while cosolute-water bonds are weaker, longer, and less linear[63,64,74]. Additional evidence for the differently structured water at the protein interface in the presence of cosolute comes from the experiments on met16. Interestingly, the heat released in the folding process is much higher than that expected for the dilution of a cosolute solution by the same amount of water released upon peptide folding, $\Delta\Gamma_w$. This would imply that waters are energetically frustrated at the macromolecular interfaces.

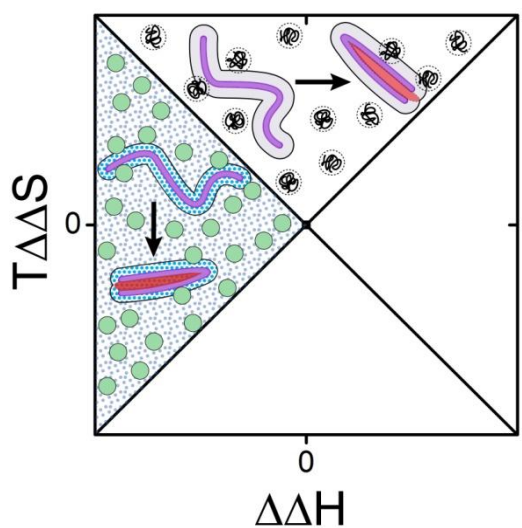

**Fig. 4. Scheme of proposed mechanisms for enthalpic or entropic cosolute stabilization of peptide folding.** The enthalpic stabilizing mechanism (sector 1, left) likely acts through solvent mediated interactions, changing the structure of solution as solvent molecules (small blue circles) are sequestered from the bulk by the cosolutes (large green circles). The entropically stabilizing mechanism (sector 2, top) acts through reduction of available volume to cosolute (black squiggles) caused by the presence of the

## 7. Concluding remarks

The abundance of information on the impact of excluded cosolutes on macromolecular processes indicates that we cannot use a single simple molecular model to collectively explain all these observations. Entropically dominated depletion forces have been extensively studied and documented, and are well explained by dominance of hard-core steric interactions. The more recently realized enthalpic depletion forces require us to explore new models for these forces. First indications implicate the solvent as an important added component that must be explicitly considered. Consideration of intermolecular forces that go beyond hard cores is also necessary. Importantly, these observations carry wide-reaching implications as to the way the cellular environment and its careful control through the cellular cosolute composition impact protein functions and interactions. We hope that additional experiments and models at the molecular level, including techniques such as electron pulse resonance and terahertz spectroscopy, will be able to detect changes in solution structure that could eventually more completely explain the cosolute effect.

## Acknowledgments

Financial support from the Israel Science Foundation (ISF Grant No. 1538/13) is gratefully acknowledged. LS is supported by the Adams Fellowship Program of the Israel Academy of Sciences and Humanities. The Fritz Haber research center is supported by the Minerva Foundation, Munich, Germany.

# Supporting Information for

# Balance of enthalpy and entropy in depletion forces

Shahar Sukenik, Liel Sapir and Daniel Harries

## Data analysis

The information shown in figure 3A,B was compiled from available literature that describes experiments involving the effect of crowders and osmolytes on protein folding and interactions. Here we detail for each reference the analyses and methodology used to obtain the changes in enthalpy and entropy for the various processes at 298K.

**Sukenik, Sapir, Politi and Harries** [1]

Our previous work examines the effects of both polymeric crowders and polyol osmolytes on the folding of a 16 amino acid peptide, met16. The measurement used circular dichroism to calculate $\Delta G$ directly, at various temperatures. The resulting van 't Hoff plot allows a determination of the value of $\Delta H$ and $\Delta S$ at 298K. Values reported here are those from the original manuscript. In addition to previously published data, the effect of TEG on met16 folding was measured as described for all other cosolutes.

**Gekko** [2]

The work examines the effect of polyol osmolytes on lysozyme stability using DSC. The experiments describe the enthalpy change upon thermal denaturation, $\Delta H_m$, evaluated by integration of the DSC curves, as well as $T_m$ of the protein under given conditions determined by the midpoint of the heat capacity peak.

To calculate $\Delta\Delta G$, $\Delta\Delta H$ and $\Delta\Delta S$ at 298K we used the following relations:

$$\Delta H = \Delta H_m - \Delta C_P(T_m - T)$$

$$\Delta G = \Delta H_m \left(1 - \frac{T}{T_m}\right) - \Delta C_P (T_m - T) + T\Delta C_P \ln\left(\frac{T}{T_m}\right)$$

$$T\Delta S = \Delta H - \Delta G$$

We subtracted the corresponding values in water from the value in the presence of cosolute to obtain the $\Delta\Delta$ values. We used $\Delta C_P$, as reported in the original manuscript, as the value of the reference in buffered water, ~7.7 kJ/mol.

**Foglia, Carullo and Del Vecchio** [3]

The work examines the effect of TMAO on the thermal stability of bovine RNase A, using DSC as well as CD spectroscopy. We derived values for the thermodynamic properties at 298K in a similar way to those described in [2], except that $\Delta C_P$ was reported in the manuscript for each cosolute concentration directly from the baseline slopes of the DSC curves for the folded and unfolded species. Values reported are compared to RNase A unfolding in 50mM HEPES at pH 7.

**Lin and Timasheff** [4]

Here, the thermal stability of RNase T1 was measured in the presence of TMAO. The measurement followed absorbance of the protein at 286 nm at various temperatures and TMAO concentrations. We converted the reported melting curves to free energy plots assuming two-state folding, and the resulting linear van 't Hoff plot gave $\Delta H$ and $T\Delta S$ at 298K. The enthalpy and entropy change upon unfolding in 30 mM phosphate buffer, pH 7 was subtracted from these values.

**Benton, Smith, Young and Pielak** [5]

This work tests the effect of ficoll, a large polymeric crowder, and its monomer sucrose on the thermal stability of chymotrypsin inhibitor 2 (CI2). The experiments used NMR to detect amide proton exchange as a way to quantify the degree of denaturation of the protein. Values for $\Delta G$ and $\Delta H$ at 310K are presented as reported in the paper, and from these we subtracted the values reported for dilute solution (50 mM sodium acetate buffer, pH 5.4).

**Wang, Sarkar, Smith, Krois and Pielak** [6]

This work reports ubiquitin thermal stability measured in the presence of polymeric crowders and non-interacting proteins using NMR amide-proton exchange. Enthalpy and entropy values compared to the protein in 50 mM sodium acetate buffer (pH 5.4) were measured as described in Benton, Smith, Young and Pielak, Biochemistry (2012). To obtain the values at 298K we fit the free energy as a function of temperature to the following equation:

$$\Delta G(T) = \Delta H(298K) + \Delta C_P(T - 298K) - T\left[\Delta S(298K) + \Delta C_P\left(\ln\frac{T}{298K}\right)\right]$$

The enthalpy and entropy values of the reference reaction in the buffer were subtracted from those in the presence of the polymers to obtain the values we report here.

**Ratnaparkhi and Varadarajan** [7]

The stability of the interaction between the N and C-terminals of RNase S was measured using ITC in the presence of several osmolytes. The values we plot for $\Delta G$, $\Delta H$ and $T\Delta S$ are as reported directly from the experiments, and are reported with respect to the reaction in 50 mM sodium acetate, and 100 mM NaCl at pH 7.

**Zhang, Wu, Chen and Liang** [8]

The reported work describes a-lactalbumin stability in the presence and absence of PEG and ficoll as measured using intrinsic Trp fluorescence. We converted the resulting sigmoidal melting curves into unfolded fraction plots assuming a two-state melting process, and subsequently converted these to van 't Hoff plots. These plots were fit with a polynomial function which was then differentiated around 298K to obtain the enthalpy and entropy of folding. The reference point was the melting reaction in 10 mM sodium phosphate buffer with 100 mM NaCl and 1 mM EGTA, pH 7.

**Milev, Bosshard, and Jelesarov** [9]

This paper describes the effects of salts and several osmolytes on the DNA binding reaction of the integrase Tn916 DNA-binding domain. The thermodynamic constants were obtained using ITC measurements, and reported here as in the original manuscript.

**Kumar, Attri and Venkatsu** [10]

The authors examine the thermal stability of α-chymotrypsin in the presence of polyol osmolytes. Values reported here were obtained in the original manuscript using DSC, and processed in the original paper to give the thermodynamic constants at 298K as described in [1].

# SPT derived free energy of dimerization at the limit of low cosolute concentration

In this section we show the equivalence of SPT and the work related to decreasing excluded volume at the low cosolute concentration limit. Let us first examine the dependence of the free energy of dimerization, $\Delta G^{dim}$, for a system composed of a solvent containing spherical crowders with radius $r_c$ at concentration $c_c$, and two macromolecule monomers with radius $r_m$, that dimerize to give a spherocylinder with equal volume to that of two monomers and the same radius. This constraint of no volume change requires that the length of the central cylindrical part of the spherocylinder is $l_d = \frac{2}{3} r_m$. The model is described schematically in Fig. 2A of the main text. Next, define $\phi$ as the volume fraction of the crowder, and denote the radius of the macromolecule scaled by that of the crowders, $R = r_m / r_c$. Finally, the scaled length for the cylindrical part is defined as $L = l_d / 2r_m$.

In order to create a cavity that allows the insertion of a macromolecule (be it monomer or dimer) into the crowder solution, a change in free energy is incurred, given by [11–13]:

$$\frac{\Delta G}{RT} = -\ln(1-\phi) + A_1\left(\frac{\phi}{1-\phi}\right) + A_2\left(\frac{\phi}{1-\phi}\right)^2 + A_3\left(\frac{\phi}{1-\phi}\right)^3,$$

where $A_1$, $A_2$ and $A_3$ are constants related to the dimensions of the inserted macromolecule. To follow how this function behaves in the limit of low cosolute concentration, we derive an expression for the change of the free energy with respect to cosolute concentration:

$$\frac{\partial(\Delta G/RT)}{\partial c_c} = \frac{\partial(\Delta G/RT)}{\partial \phi}\frac{\partial \phi}{\partial c_c} = \left[\frac{1}{1-\phi} + A_1\left(\frac{1}{1-\phi}\right)^2 + 2A_2\phi\left(\frac{1}{1-\phi}\right)^3 - 3A_3\phi^2\left(\frac{1}{1-\phi}\right)^4\right]\frac{\partial \phi}{\partial c_c}$$

In the limit $\phi \to 0$ we can neglect the terms involving $A_2$ and $A_3$. The expression then reduces to

$$\lim_{\phi \to 0}\frac{\partial(\Delta G/RT)}{\partial c_c} = (1+A_1)\left(\frac{\partial \phi}{\partial c_c}\right),$$

with $A_1 = R^3 + 3R^2 + 3R + 1.5LR(R^2 + 2R + 1)$. For the dimerization process we consider $\Delta\Delta G = \Delta G^{dim} - 2\Delta G^{mono}$, where *dim* and *mono* superscripts indicate the dimer and the monomer molecules, respectively. This free energy change then results in the final expression:

$$\left.\frac{\partial(\Delta\Delta G/RT)}{\partial c_c}\right|_{c_c \to 0} = -\left(1+\frac{r_m}{r_c}\right)^2\frac{\partial \phi}{\partial c_c} = \Delta V_{ex}^{dim}$$

Where we have used the relation

$$\frac{\partial \phi}{\partial c_c} = \frac{4}{3}\pi r_c^3$$